\documentclass[amsmat,amssymb,amsfonts,aps,prb,twocolumn,showpacs]{revtex4}

\usepackage{graphicx}
\usepackage{dcolumn}
\usepackage{bm}

\begin{document}

\title{Vacancy induced magnetism in graphene and graphene ribbons}

%\preprint{1}
\author{J.\ J.\ Palacios}
\affiliation{Departamento de F\'\i sica Aplicada, Universidad
de Alicante, San Vicente del Raspeig, E-03690 Alicante, Spain.}
\author{J. Fern\'andez-Rossier}
\affiliation{Departamento de F\'\i sica Aplicada, Universidad de Alicante,
San Vicente del Raspeig, E-03690 Alicante, Spain.}
\author{L. Brey}
\affiliation{Instituto de Ciencia de Materiales de Madrid,
 Consejo Superior de Investigaciones Cient\'\i ficas,
E-28049 Cantoblanco, Spain}

\date{\today}

\begin{abstract}

We address the electronic structure and magnetic properties of vacancies and
voids both in graphene and graphene ribbons. Using a mean field Hubbard model,
we study the appearance of magnetic textures associated to removing a single atom
(vacancy) and multiple adjacent atoms (voids) as well as the magnetic interactions between
them. A simple set of rules, based upon Lieb theorem,  link the atomic
structure and the spatial arrangement of the defects to  the emerging magnetic order. The total spin $S$
of a given defect depends on its sublattice imbalance,  but some defects
with $S=0$ can still have local magnetic moments. The sublattice imbalance also
determines whether the defects interact ferromagnetically or 
antiferromagnetically with one another and the  range of these magnetic
interactions is studied in some simple cases.  We find that in semiconducting
armchair ribbons and two-dimensional graphene without global sublattice imbalance there is maximum defect
density above which local magnetization disappears.  
%The critical inter-defect distance
%increases as the gap of the ribbon decreases.  
Interestingly, the electronic properties of
semiconducting graphene ribbons with uncoupled local moments are very similar to those of
diluted magnetic semiconductors, presenting giant Zeeman splitting. 
%Interestingly, vacancy induced midgap states resemble  Su-Schrieffer-Heeger
%states in polyacetilene and display spin-charge separation in striking contrast
%to acceptors and donors in semiconductors. 

\end{abstract}

%\pacs{PACS numbers: }

\maketitle

\section{introduction}
Magnetic order occurs, in most instances, in materials with partially filled $d$
or  $f$  shells. There is, however, 
a recent awareness that  the possibility of
magnetic order can  also occur  in  
materials without open $d$ or $f$ 
shells\cite{Coey05,Esquinazi07,JFR07,gallego,pantelides}.  
Experimental evidence of 
this new type of magnetism
has been found in thin films of certain  oxides  
(HfO$_2$, ZnO, TiO$_2$)\cite{Coey05}, as well as  irradiated
graphite\cite{Esquinazi07}, and  thiol-capped 
gold nanoparticles \cite{Crespo,Hori}.

Although  more experimental work is probably necessary 
to confirm and understand magnetism in 
these systems,  there are at least two scenarios for which theory
provides a mechanism for the appearence of magnetism without $d$ or $f$ open
shells. On one side, in some lattices
intrinsic lattice
defects like vacancies lead to the formation local magnetic moments, a
preliminary condition for the existence of magnetic order. 
This is the
case in graphite\cite{Lethinen04},  
graphene\cite{Vozmediano05,Kumazaki07,Yazyev07} 
and II-VI semiconductors\cite{Lannoo08}. On the other side, it has been
recently found that    clusters with specific shapes, like triangular graphene islands\cite{JFR07} 
or icosahedral\cite{gallego,pantelides} gold clusters,  have large
degeneracies at the Fermi energy in their single particle  spectra. 
These degeneracies are related to the symmetry
of the nanostructure and, in words of Luo {\em et al.}\cite{pantelides},
they behave like 'superatoms', with magnetic ground states that comply with
atomic-like Hund's rules. 

Importantly, both vacancy induced\cite{Kumazaki07,Yazyev07} 
and 'superatomic' magnetism\cite{JFR07} occur in graphene
structures and, as we show here, have the same origin.
 In this work we present  extensive numerical work 
to understand vacancy induced magnetism in graphene and graphene ribbons 
and we analyze our results in the context of  a broader theoretical 
framework  that unifies 
superatomic\cite{JFR07} and vacancy induced magnetism\cite{Kumazaki07,Yazyev07}
  in graphene. In part our motivation stems from the recently shown possibility of 
fabricating high mobility graphene based field effect 
transistors\cite{Novoselov04,Bunch05,Geim05,Kim05,Science06,RoomQHE} which has created enormous
interest in graphene based electronics\cite{Natmat}. Additional possibilities arise
from the fabrication of semiconducting
graphene ribbons\cite{Avouris,kim07,Pablo07} and graphene nanoislands\cite{Natmat,billiard} 
with top-down techniques  as well as the growth of graphene islands 
with bottom-up techniques\cite{Wu07,Amadeo}. 
Magnetic order  in  patterned or nanostructured graphene would
bring  up new opportunities of research in spintronics.

Graphene honeycomb structure is a bipartite lattice, formed by two  
interpenetrating triangular sublattices, A and B, such that the
first neighbors of an atom A belong to the sublattice B and viceversa\cite{book}.
The low-energy electronic structure of graphene can be described by a
single-orbital ($p_{\rm z}$) nearest-neighbor hopping  Hamiltonian\cite{Wallace47,book}. 
This model correctly describes two dimensional graphene
as a zero gap semiconductor with linear bands around the Fermi energy.
The single particle spectrum of a nearest-neighbor tight-binding model in a
bipartite lattice  has  particle-hole symmetry\cite{Abrahams,Naumis07}. 

The magnetic properties of both graphene-based nanostructures and defective
graphene are intimately related
to the appearance of midgap states and how they are affected by
electron-electron interactions.
The existence of
zero-energy states in disordered bipartite lattices was proved by
Inui et al.\cite{Abrahams}.  
Within the first-neighbor tight-binding model, a sufficient 
condition\cite{Abrahams,JFR07} for the existence of midgap states is the
existence of a finite sublattice imbalance, $N_I\equiv N_A-N_B$,  where $N_A$
and $N_B$ are  the number of atoms belonging to each sublattice  or missing from
each sublattice in an otherwise  perfect system.
Thus, whereas ideal graphene has
$N_I=0$ and no midgap states, both  defective graphene 
and some graphene islands, such as
triangles, can present finite sublattice imbalance and $|N_I|$ midgap states.
The result of Inui et al.\cite{Abrahams} has been  used in a recent
work on disorder in graphene by Pereira {\em et al.}\cite{Pereira}. 
Incidentally, the existence of zero-energy or midgap
states in uncompensated graphene structures was
known long ago in the context of chemical studies of hydrocarbons  as the
Longuet-Higgis conjecture\cite{longuet-higgins:265}.  

Because of the particle-hole symmetry, midgap
states are  half filled for neutral graphene and the appearence
of magnetic moments is expected in analogy with Hund's rule in atomic
magnetism. The Hubbard model extends the single-particle tight-binding model
including the effect of Coulomb repulsion between two electrons in the same
atomic site. Importantly, a theorem by Lieb, valid for the exact ground state of the  Hubbard model
and neutral bipartite lattices\cite{Lieb89} states  
that the total spin $S$ of the  ground state is given by $2S=|N_A-N_B|=|N_I|$.
Lieb's theorem provides a rigorous connection between vacancies in the graphene
lattice and  the emergence of magnetism. As as result, sublattice unbalanced neutral graphene will always present a
finite total magnetic moment.

Although Lieb's theorem provides the total spin of the ground state, it 
does not say much about the actual local magnetic order or spin texture.
For instance, $S=0$  does not preclude the existence of local magnetic moments 
coupled antiferromagnetically or presenting
compensated ferrimagnetic order. The most notorious example of
compensated ferrimagnetic order can be found in zigzag
ribbons\cite{Waka98,son:216803,Cohen-Nature,pisani:064418,JFR-preprint},  
where each edge presents ferromagnetic order antiparallel 
to each other for a total vanishing magnetic moment.
Other examples can be found in hexagonal
graphene islands, where,  beyond a critical size, contiguous sides alternate the
direction of the ferromagnetically ordered magnetic moments\cite{JFR07}. 
%Vacancy pairs in bulk
%graphene\cite{Yazyev07} couple antiferromagnetically if they belong
%to different sublattices.

The rest of this paper is organized as follows.
In Sec. \ref{method} we review the single orbital Hubbard model and the different methodologies
 used to describe the electronic structure of defective graphene and graphene ribbons.
The underlying non-interacting spectrum and associated magnetic textures can
 be anticipated following some basic rules which are presented
 in Sec. \ref{rules}.  We illustrate the
validity of the rules by numerical calculations 
in the case of semiconducting armchair ribbons (Sec.
\ref{single})  with vacancies, voids, or notches, both in the non-interacting
(Sec. \ref{single}) and interacting (\ref{U}) cases.
The results for bulk graphene are discussed in
Sec. (\ref{bulk}). Summary and conclusions are presented in Sec.  \ref{conclusions}.

\section{Methodology}                                          
\label{method}

We consider the low-energy physics that takes place in the subspace expanded only by the single
$p_z$ orbital  (the one perpendicular to the graphene plane). Next-to-near neighbor hopping is neglected and
the electron-electron interactions are included locally in the form of an on-site repulsion or Hubbard model. When
the interactions are turned off this reduces to the
widespread one orbital tight-binding model\cite{Wallace47,Nakada96,Brey06,Munoz06}. 
The Hubbard term is treated in a mean field approximation\cite{Waka98,JFR07}.
Comparison between the results so obtained and density functional theory (DFT) calculations yield very good
agreement for two-dimensional graphene\cite{Ordejon}, carbon nanotubes\cite{Ordejon,Pilar}, 
zigzag\cite{Cohen-Nature,JFR-preprint} and armchair graphene ribbons\cite{son:216803}, as well as
 graphene islands\cite{JFR07}. 
 
We model vacancies and  voids in perfect
graphene or graphene ribbons by removing atoms, actually, by 
removing the representing $p_z$ orbitals in the tight-binding
model. This results in a reduction of the coordination of the  atoms adjacent to the missing atoms.
We ignore the  lattice distorsion and we assume that the on-site
energy is the same for edge and bulk atoms.
The single-orbital hamiltonian implicitly assumes full hydrogen passivation of the $sp_2$ dangling bonds of the atoms
without full coordination.  This assumption, which might not be completely
realistic in the case of actual vacancies\cite{Lethinen04}, does not invalidate our model; 
for we can consider an alternative physical realization: The
chemisorption of a hydrogen atom on top of a bulk graphene atom\cite{Yazyev07} effectively removes a $p_z$ orbital
from the low-energy hamiltonian. In our
one-orbital model there is no difference between these two scenarios. 

DFT calculations on graphene ribbons\cite{son:216803,pisani:064418}, graphene
islands\cite{JFR07,Scuseria,Kaxiras08}, and bulk graphene with
vacancies\cite{Yazyev07} have shown that the results follow the predictions
of  Lieb's theorem, even though DFT calculations go beyond the first-neighbor
hopping, short-range interaction Hubbard model on which the theorem is based. In
other words, second neighbour hopping  and inter-site Coulomb
repulsion, present in the DFT calculations, do not modify the relation
between lattice imbalance and total spin of the ground state warranted for the
Hubbard model for which these couplings are absent. From this point of view
these couplings are irrelevant. It is thus justified to consider the following mean-field
hamiltonian:
\begin{equation}
\label{H-F}
 H=H_0+ U\sum_i (n_{i\uparrow}\langle n_{i\downarrow}\rangle +
 n_{i\downarrow}\langle n_{i\uparrow}\rangle) - U\sum_i \langle n_{i\downarrow}\rangle
 \langle n_{i\uparrow}\rangle ,
 \end{equation}
where $i$ runs over all lattice sites and the non-interacting hamiltonian reads
\begin{equation}
H_0=\sum_{i,j} t(c_i^\dagger c_j+c_j^\dagger c_i),
\end{equation}
where the sum runs over nearest-neiborgh lattice sites $i,j$ and $t=2.5 $ eV.
Without loss of generality, we have set the diagonal terms of the Hamiltoninan to zero. For neutral graphene
we can rewrite the mean-field  hamiltonian (up to a constant) as the sum of two terms:
\begin{equation}
H=H_0 + \frac{U}{2}\sum_{i} n_i \langle n_i \rangle - U \sum_i 2 m_i \langle m_i\rangle
\end{equation}
where $m_i=\frac{1}{2}\left(n_{i\uparrow}-n_{i\downarrow}\right)$ and $n_i=n_{i\uparrow}+n_{i\downarrow}$.
The second term in $H$ represents the non-trivial contribution of interactions.

The calculations have been performed considering three different types of
boundary conditions. For the evaluation of the non-interacting density of states
(DOS) in ribbons we compute the Green's function projected on  the region where the defects
are located. The perfect regions of the ribbon away from the defects are
included in the Green's function by means of a self-energy. We refer the reader
to  Ref. \onlinecite{Munoz06} for more details on this methodology.
When the interactions are turned on ($U\ne 0$) we consider ribbons with periodic
boundary conditions in one direction or, for bulk graphene, periodic boundary
conditions in both directions. More details will be given in the respective sections.

\section{Basic rules}
\label{rules}

%first-neighbour model. 
%The results can be rationalized using the ITA theorem\cite{Abrahams}.
%%This theorem is valid for tight-binding
%%models in a bipartite lattice with arbitrary 
%%first-neibhour hopping and no diagonal disorder.
%The theorem states that for a given sublattice imbalance $N_I$ there are 
%$|N_I|$ zero energy states. On top of that,
%the zero energy states are sublattice resolved and they have no weight on the
%minority sublattice. These mid-gap states are central to our work
%since are responsible for most of the electronic properties discussed below.  
%Since perfect graphene has $N_I=0$, the removal 
%of an odd number of atoms  results in a lattice imbalance, 
%whereas the removal of an even number of atoms could result in either
%compensated or uncompensated structures.

\begin{figure}
[hbt]
\includegraphics[width=\linewidth]{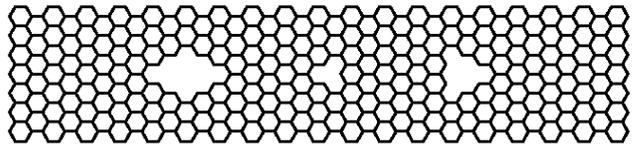}
\caption{ \label{voids} Examples of voids  with different sublattice imbalance
in the middle of a graphene ribbon. From left to right the
associated imbalance charge is $N_I=0,-1$ (vacancy), and 2.}
\end{figure}

In this section we provide some general rules to understand the appearance of midgap
states and magnetic textures due to single-atom vancacies or, more
generally, voids in an otherwise sublattice balanced graphene
structure, for instance,  an infinite defect-free semiconducting armchair ribbon. 
%For the ensuing discussion we will consider voids %Some aspects of the  ensuing
%discussion need to be revised in the case of gapless graphene systems (see Sec.
%\ref{gapless}) since they have a profound effect on the conclusions presented in
%this section.
A void in graphene  can be partially characterized by the number
and type, A or B, of atoms removed from the otherwise perfect structure. We will label voids as one
would do for chemical compounds, A$_{N_A}$B$_{N_B}$. Voids
will be unbalanced when the are created by removing $N_A$ and $N_B$ atoms such
that $N_I=N_A-N_B \ne 0$ 
The sublattice imbalance $N_I$, which can be either positive or negative, can be interpreted as an imbalance ``charge''.
This quantity is
central to our discussion, although the exact formula of the void is also
important since it gives an idea of the size and  shape of the void. In the
case of ribbons the voids can be close to the edges, becoming thus notches. (For
the most part we will refer to voids in the bulk of the ribbon, but the
conclusions apply equally to the case of notches on the edges.)
For a single unbalanced void characterized by $N_I$,  $|N_I|$ zero-energy
states appear in the non-interacting spectrum with weight on only one
sublattice\cite{Abrahams}. When the graphene structure presents a gap $E_{\rm g}$, as in armchair ribbons,  
these states
are  normalizable and localized around the void (in contrast to two-dimensional graphene
where the zero-energy states are not normalizable\cite{pereira06}). Figure
\ref{voids} shows various examples of voids in a ribbon with
different sublattice imbalance $N_I$. 

Let us consider now two voids, characterized {\em locally} by $N_I(1)$ and $N_I(2)$,
sufficiently separated so that they do not affect each other.
According to the result of Inui et al.\cite{Abrahams}, the single particle spectrum
has, at least,
\begin{equation}
\label{ITA}
N_Z^{\rm min}= |N_I(1)+N_I(2)|
 \end{equation}
  midgap states. The important question is what happens when the distance between
them is such that they do affect each other. 
 If $N_I(1)$ and $N_I(2)$ have
the same sign, there are $|N_I(1)|+|N_I(2)|$ midgap
states, regardless of the distance. 
If they have different signs, e.g., $N_I(1)+N_I(2)=0$, Eq. \ref{ITA}
apparently warrants the annihilation of midgap states.
Within the non-interacting model  midgap states are
100 percent sublattice polarized.  The non-interacting Hamiltonian has finite
matrix elements between states that have weight on different sublattices. Hence,
the mechanism for midgap state annihilation is hybridization of midgap states
localized in different sublattices. This annihilation occurs as
bonding-antibonding pairs of midgap states form, resulting in  a shift in their
energy  and in a loss of the sublattice polarization.  For large distances, however, this 
annihilation does not occur.

A well understood related example occurs in zigzag
ribbons\cite{Nakada96,Brey06}. The edge of
a zigzag ribbon has a local sublattice imbalance. If the top edge belongs to the
A sublattice, the bottom edge belongs to the B sublattice. States fully
localized in the edge have zero energy and are localized in a single sublattice.
To the extent that  states mostly localized on the top edge penetrate into the ribbon, 
they hybridize with  states mostly localized on the bottom edge. This mixing
results in a bonding-antibonding splitting that takes these states away from the
Dirac energy. In the case of zigzag ribbons, the degree of localization in the
unit cell depends on the wavevector. States close to the zone boundary are very
localized in the edges and have energy very close to zero\cite{Nakada96,Brey06}.
The localization
decreases as the wavevector departs from the boundary, resulting in the
hybridization and the departure from zero energy\cite{Nakada96}.

%When we consider two  voids that, alone, would create 
%$N_1$ midgap states in sublattice $A$ and $N_2$ midgap states in sublattice $B$,
%the number of midgap states with energy very close to zero
%can be as large as 
%\begin{equation}
%N_Z^{max}=|N_1|+|N_2|
%\end{equation}
In the general case one can conclude that the minimum number of zero-energy states will be given by
\begin{equation}
N_Z^{min}=\sum_{\alpha,\beta}|N_I(\alpha)+N_I(\beta)|,
\end{equation}
where the integer indeces $\alpha$  and $\beta$ run over voids with the same imbalance sign, respectively.  
In practice, within an arbitrarily small energy interval $|E|\rightarrow 0$, 
the number of zero-energy states can be as large as 
\begin{equation}
N_Z^{max}=\sum_{\alpha}|N_I(\alpha)|+\sum_{\beta}|N_I(\beta)|.
\end{equation}
In the general case, $N_Z$ will be a number between $N_Z^{min}$ and $N_Z^{max}$.

When electron-electron interactions are turned on, at least locally in the form
of a Hubbard-type interaction, Lieb's theorem guarantees that
$|N_I|=2S$ for neutral graphene. 
The theorem, however, does not exclude the possibility of spin-symmetry broken local magnetic order
when $S=0$ or small. For instance, two or more voids with local sublattice imbalances that cancel out 
the total imbalance can still retain their 
local magnetic order when they are not in proximity. When the imbalance of the void is zero but the size
is large an internal ferrimagnetic order cannot be discarded either.  
In general, calculations will be necessary to ascertain the spin texture in these situations.
A few conclusions, however, can be reached without actually performing any calculations.
One can distinguish four cases:

\begin{enumerate}
\item $N_Z^{min} = N_Z = N_Z^{max}$: In this case all the voids are of the 
same sign. The coupling between them
is always ferromagnetic and the spin of the ground state is $2S=N_Z$.
The splitting with smaller spin states will depend on the inter-void coupling.

\item $N_Z^{min} = N_Z < N_Z^{max}$:  In this case all the voids of different 
sign are in proximity and interact, yielding a $2S=N_Z^{min}$ state. 
Calculations will be necessary to ascertain the spin texture in these situations.

\item $N_Z^{min} < N_Z = N_Z^{max}$:  In this case all the voids of different
type are separated and uncoupled. The ground state has $2S=N_Z^{min}$, but 
the spin-flip gap is negligible since there are uncoupled magnetic moments.

\item $N_Z^{min} < N_Z < N_Z^{max}$:  In this case there are voids of different
sign, but some  of them are uncoupled and some not.  This is the most general case.
The ground state has $2S=N_Z^{min}$, but, as in the previous case,
the spin-flip gap is negligible since there are uncoupled magnetic moments.
Calculations will be necessary to ascertain the spin texture in these situations as well.

\end{enumerate}

\section{Defects in semiconducting graphene ribbons: Non-interacting theory}                                  
\label{single}
In this section we study the electronic structure of defective graphene ribbons within the
non-interacting tight-binding model.
The results are obtained using the cluster embedded
method described in Ref. \onlinecite{Munoz06}.  In this methodology
a finite portion of the ribbon containing the defects 
is attached to two semi-infinite perfect ribbons of the same
width and compute the DOS of the defective region. 
In the defect free case we obtain a gap in the DOS.
We  consider armchair graphene ribbons of width $W=N_y a$, 
where $N_y$ is an integer number and $a=2.42\:$ \AA is the graphene lattice
parameter. We only consider values of $W$ such that, within  the
first-neighbor  tight-binding model, the ribbon is semiconducting\cite{Nakada96,Brey06}. 
This happens  if $N_y+1$ is not a multiple of three.  More
realistic  calculations\cite{son:216803} predict that, because of lattice
distorsion of the edge atoms, even ribbons with $N_y+1=3m$, where $m$ is an
integer, are semiconductors.  Semiconducting graphene ribbons attract interest
due to possible applications in nanoelectronics\cite{Avouris,kim07,JFR-ribbonAC-07,Munoz08}. 
As in the case of Si based 
semiconductors, their electronic structure might be strongly influenced by
impurities.  Here we study the effect of vacancies and voids, which are expected to act as
neutral impurities.

\subsection{Single void}
\begin{figure}
[hbt]
\includegraphics[width=\linewidth]{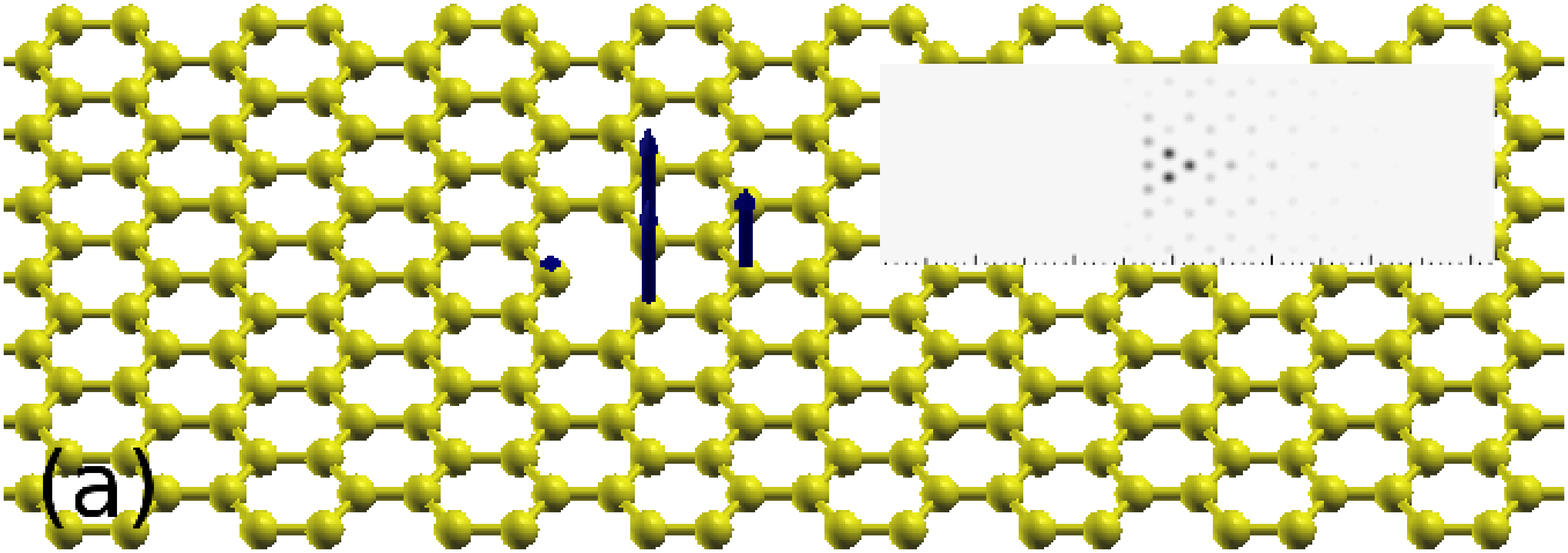}
\includegraphics[width=\linewidth]{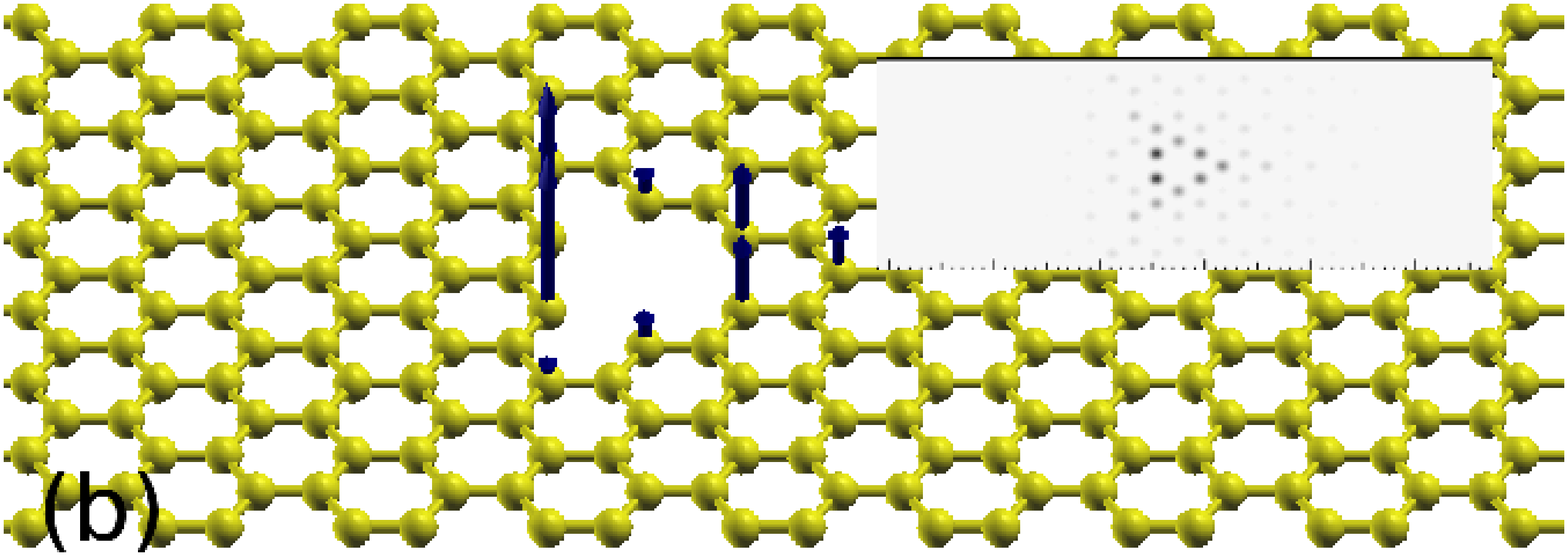}
\caption{ \label{figsingle} (Color online) (a) Magnetic moments on lattice sites around a
single vacancy. Inset: Probability density of the  zero-enery state built with
gaussian functions on lattice sites.  (b) Same as in (a) but for a
triangular void with $N_I=2$. Inset: Same as in (b) but summing over   the two
zero-energy states.} 
\end{figure}

The simplest defective structure is a perfect 
semiconducting graphene ribbon from which a single atom, A or B, is removed.
In agreement to Eq. \ref{ITA}, a zero energy state appears in
the DOS. For neutral graphene, this state is half filled. In other words, an
spin unpaired electron  occupies the midgap state. 
The local density of states at zero
energy, which is nothing but the modulus square of the wave function associated with 
the zero energy state, is
shown in the inset of Fig. \ref{figsingle}(a). The state is localized in 
the neighborhood of   the vacancy. The shape of the midgap state is also
peculiar: It has a clear directionality. Monoatomic
vacancies have the shape of a triangle. 
Two vertices of the triangle point towards
the edges of the ribbon, whereas the lateral vertex 
 can point downstream or upstream
along the ribbon.  The midgap  state is peaked around the lateral vertex.
Hence, midgap states have a strong directional character.

Importantly, the {\em integrated} charge, including both mid-gap and band
states below the Fermi energy, yields a homogeneous charge distribution: there
is one electron per atom in every atom, even in the presence of the vacancy.
Hence, the localized midgap state does not imply charge localization, 
yet  there is a finite spin density. The total spin $S$ of the neutral
graphene with one vacancy is $1/2$ and the spin density does show a
non-homogeneous texture, as shown in  Fig. \ref{figsingle}(a). 
Hence, the region of the
material around the defect has no local charge but has local spin. 
%Globally,
%the charge is zero but the spin is $1/2$.  
%This spin-charge separation is  very similar to that of
%Su-Schrieffer-Heeger  (SSH) states localized in the domain walls in
%polyacetilene (\cite{SSH}). 

Our next step is to consider larger voids. In Fig. \ref{figsingle}(b) we show
the results for a void with $N_I=2$, obtained by removing four atoms (A$_3$B$_1$). In agreement
to Eq. \ref{ITA}, there are two midgap states (per spin). Their local density of
states is shown in the inset of Fig. \ref{figsingle}(b). As in the case of
monoatomic vacancy, two electrons occupy the midgap states. 
The {\em integrated} local charge is also homogeneous: One electron per site.
 Within
the framework of the non-interacting model we cannot discriminate between the
$S=0$ or the $S=1$. As we discuss below, when Hubbard repulsion is turned on
Lieb's theorem warrants that the spin of this structure is $S=1$. In Fig.
\ref{figsingle}(b) we show the magnetization density, calculated including the
interactions, as discussed below.  As in the case of a sigle missing atom, there
is a magnetic texture with $S=1$ which is localized
in a region without localization of extra charge.  
Triangular voids maximize $|N_I|$ while removing the minimum number of atoms. Larger ones ($|N_I|>2$)
exhibit similar features to the ones already discussed. More complicated voids
with zig-zag edges such as hexagons or rombhi, which have $N_I=0$,
can still present quasi zero-energy states if they are sufficiently big and therefore
might exhibit spin textures as discussed below.

\subsection{Two voids}

As a step towards understanding the electronic structure of graphene with a
finite density of defects, we  first consider the electronic structure of two
voids with the same absolute value of $N_I$. 
Each void has a well defined sublattice imbalance number $N_I$ when apart, which
can be positive  or negative.  If the sublattice imbalance of the two voids
has the same sign, the global structure has twice as many zero energy
states as the separated defects.  The non-interacting hamiltonian does not
couple sites on the same sub-lattices so that the zero-energy  states associated
with  two vacancies with sublattice  imbalance  of  the same sign cannot
interact, regardless of the distance separating them, i.e., the non-interacting
DOS  will always present a  delta function at zero
energy that can accomodate $2\times |N_I|$ electrons per spin channel. 

When the imbalance numbers are of different sign they cancel out each other.
When the defects are far away from each other their
local electronic structure is expected to be the same as that of a single defect: 
Midgap states localized in a single sublattice around the
missing atoms.  As the defects become closer, the single particle Hamiltonian,
which couples atoms of different sublattices, will hybridize the otherwise
sublattice polarized midgap states, which will result in bonding and antibonding
pairs away from zero energy. The localization length of the single defect states sets
the length scale at which this hybridization occurs. 

\begin{figure}
[hbt]
\includegraphics[width=\linewidth]{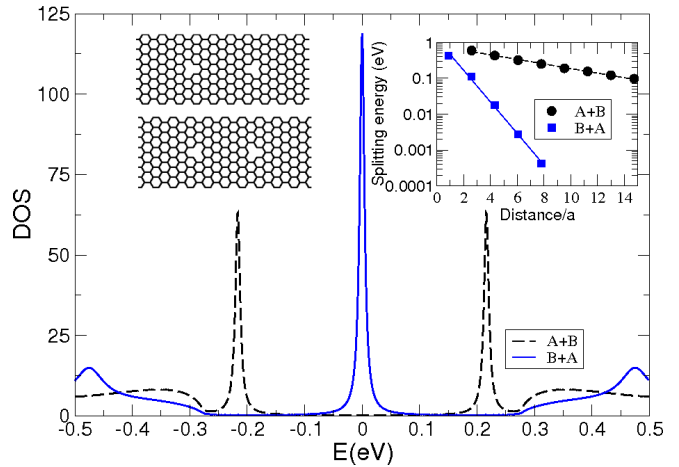}
\caption{ \label{A-B.B-A.dos} (Color online) Density of states near the Dirac point for
an armchair ribbon of $W=7a$ with two vacancies presenting charge of different sign and same modulus ($N_I=\pm 1$). 
Solid lines correspond to the B+A case (left lower inset) and dashed lines correspond to the A+B case (left upper inset). 
A finite broadening has been added for visibility's sake of the delta functions.
The finite, but small, energy splitting in the former case is not visible for this broadening.
Right inset: Bonding-antibonding energy splitting as a function 
of the distance between vacancies for the two different spatial orderings.}
\end{figure}

Our numerical calculations confirm this scenario. For a given width,  the
hybridization depends on the distance and, given the directional character of
the midgap states in ribbons, on the relative orientation. 
%Tail to tail (B+A) versus
%head to head (A+B).  
In Fig. \ref{A-B.B-A.dos} we show the DOS for a system with
two monoatomic vacancies A and B ($N_I=\pm 1$, respectively).   They are
aligned along the ribbon axis and  placed at a distance of $6.35a$
away from each other for the two possible spatial orderings, A+B (head
to head) and B+A (tail to tail) as shown in the left insets. Due to the high directional character of the
associated zero-energy states,  the coupling is not invariant against the
interchange of positions  and the zero-energy states hybridize differently,
depending on the spatial ordering.  In one case the two-fold zero-energy peak
clearly splits into two above and below the Fermi energy. In the other the
splitting is  much smaller (not visible in this scale). For one relative
orientation the wave functions overlap and the degeneracy is strongly removed.
For the other the  wave functions do not couple at this distance and the
degeneracy is practically unaffected. 
In the right inset of Fig. \ref{A-B.B-A.dos} we  show a logarithmic plot of the energy
splitting as a function of the distance for the two cases. The splitting decays
exponentially in both, reflecting the localized character of the zero-energy
states.

\begin{figure}
[hbt]
\includegraphics[width=\linewidth]{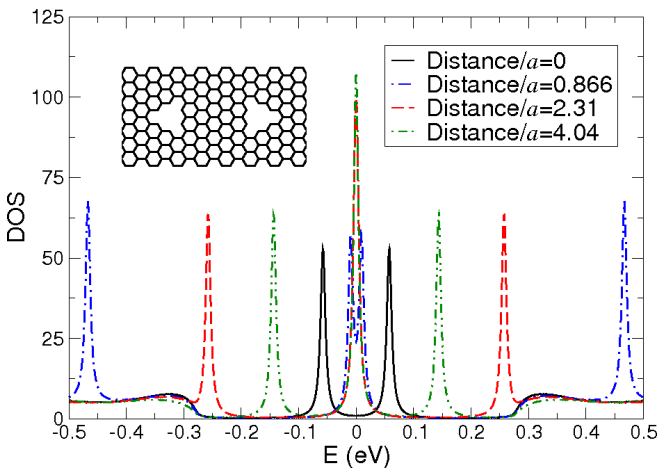}
\caption{ \label{2A-2B} (Color online) DOS projected on the vicinity of
 two triangular voids placed along the axis of a
 semiconducting ribbon for different relative distances. The imbalance charges are the same, but
differ in sign ($N_I=\pm 2$).  }
\end{figure}

We now consider the case of pairs of defects with larger sublattice imbalance.
Figure \ref{2A-2B} shows the DOS for two triangular  voids
characterized by $N_I=+2$ and $N_I=-2$ (A$_3$B$_1$ and A$_1$B$_3$, respectively) 
at different distances. We have selected
only one possible ordering in this case (tail to tail).  According to these sublattice
imbalances each void has associated two localized states. These two states also
present a strong directional character, but is different for the two.
This can be inferred from the two different bonding-antiboding splitting energies
for a given distance seen in Fig. \ref{2A-2B}. We note that,  even when the voids
approach each other, the splitting associated with one of the localized states
remains small,  still being practically zero for small distances
Only in the extreme limit of zero distance when the two voids 
merge into a single one with sublattice imbalance $N_I=0$ (A$_4$B$_4$)
there are no zero-energy states.

As the sublattice imbalance of the merging  voids becomes bigger and these condense
into even bigger $N_I=0$ voids, the number of states that appear in a vicinity 
of zero $|E| \rightarrow 0$ increases with the charge of these. 
Since the appearance of magnetic order relies on the existence of zero-energy
states, large voids with $N_I=0$ can still present ferrimagnetic order, the only
condition being that they are formed out of voids with large sublattice imbalance.
In other words, their contours must  present sufficiently long zig-zag sections.
This limits the possible shapes of these voids to, e.g., rombohedral (see Fig.
\ref{rombo}) or hexagonal forms.  This conclusion is no different from that
reached on graphene hexagonal  islands\cite{JFR07} 
or finite
length ribbons\cite{Scuseria}, where calculations have revealed compensated
ferrimagnetic order developing along the edge beyond a certain critical size.

Finally, in order to stress the fact that there is nothing in the previous
discussion specific to voids in the bulk of the ribbon, we compute the
non-interacting DOS for an A$_6$B$_4$ void plus an AB$_2$ void placed on the edges (i.e.,
notches) with $N_I=2$ and $N_I=-1$, respectively.
Removing just one atom to create a notch with $N_I=-1$ would
have given the same charge as the AB$_2$ defect, but it would  be chemically very unstable and we
ignore that possibility.  The notches are located on opposite edges,
although the results apply the same for notches on
the same edge.  A single doubly-degenerate state appears at zero energy for the
$N_I=2$ notch. When the second notch is added in close proximity
only a single zero-energy state remains,  according to the total sublattice imbalance
of the system $N_I=2-1=1$.

\begin{figure}
[hbt]
\includegraphics[width=\linewidth]{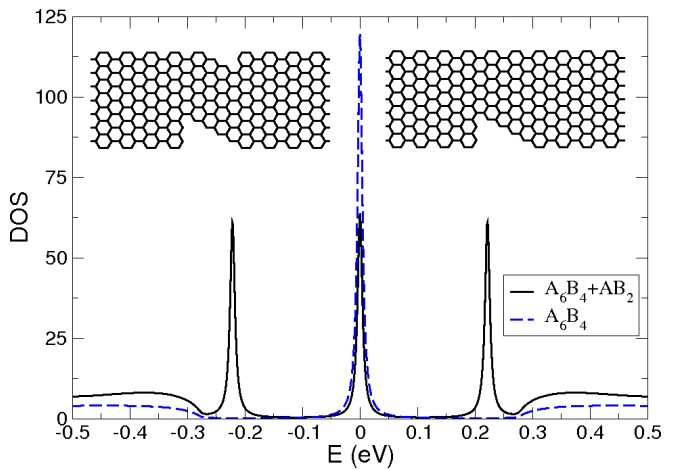}
\caption{ \label{2B}(Color online) DOS for a single notch with imbalance charge $N_I=2$
(blue dashed line, right inset). The same notch with an additional notch nearby of charge $N_I=-1$
(black solid line, left inset).}
\end{figure}

In summary, defective structures with sublattice imbalance result in half-filled
midgap states that are expected to yield magnetic moments when interactions are
turned on. Structures with global sublattice balance can still present midgap
states and be prone to developing local magnetic order, at least in two
situations: Distant defects with $N_I$ of opposite sign and large voids with
sufficiently long zigzag edges.  

\section{Defects in semiconducting ribbons: Interaction effects}
\label{U}
In this section we verify whether the physical picture anticipated from 
the non-interacting model remains true when the on-site Hubbard
repulsions are included. As shown in previous section,
%, midgap states with
unpaired spins appear in  sublattice imbalanced structures.
%These states display
%spin-charge separation, analogous to that of SSH midgap states\cite{SSH}. 
When $|N_I|>1$,  
the non-interacting model predicts that a shell of $|N_I|$ degenerate mid-gap
states is half-filled. The maximization of the spin is expected when Coulomb
repulsions are turned on, in the spirit of Hund's rule. 
%This is done using a Hubbard model in a mean field
%approximation described in the methodology section.  
At half-filling, the exact
ground state of the Hubbard model for a bipartite lattice such as that of graphene
satisfies Lieb's theorem\cite{Lieb89} which relates the 
sublattice imbalance and the ground state total spin:
 $2S= |N_I|$. For unbalanced structures this inmediately confirms the
 Hund's rule scenario. 
In the case of balanced structures the ground state spin must be zero,
but  this could happen with 
local moments, as it happens on the edges of infinite graphene ribbons. 

The numerical calculations of this section are done with a unit cell of width
$W$, length $L=N_x\frac{\sqrt{3}}{4}a$, where $N_x$ is the number of carbon atoms
along an armchair chain, and with periodic boundary conditions along the $x$
direction to  avoid spureous zigzag edges. We consider unit cells as long as 25
nm  and the typical number of atoms in a self-consistent calculation is
1000. Importantly, our mean field results have the same relation between
the sublattice imbalance and ground state spin than the exact state, as
predicted by Lieb theorem.

\subsection{Single void with $U\neq 0$}

\begin{figure}
[hbt]
\includegraphics[width=\linewidth]{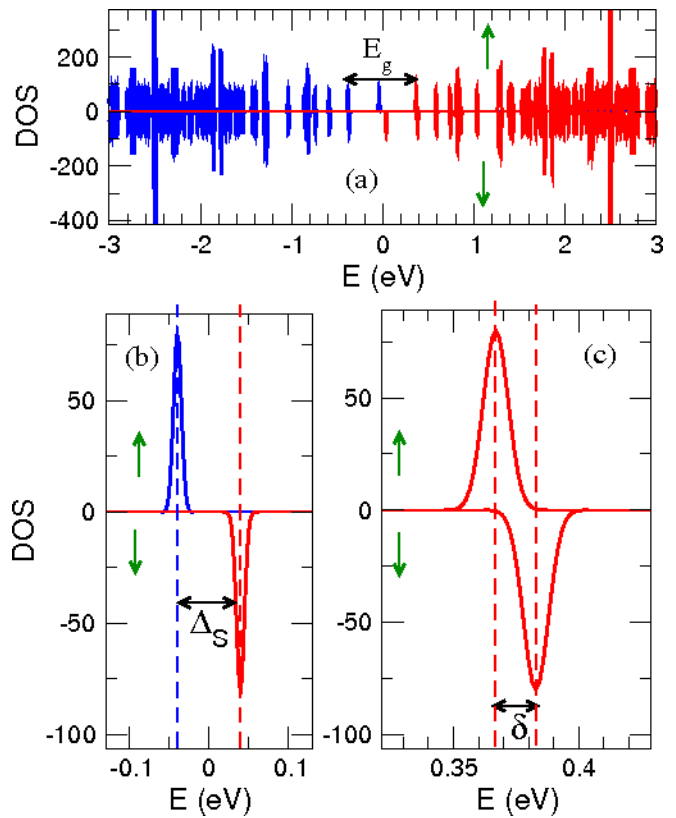}
\caption{ \label{DOS}  (Color online)
 (a) The spin resolved DOS for a ribbon with $W=7a$ and $U=2$eV with one vacancy. 
Spin $\uparrow$ ($\downarrow$)
is plotted as a positive (negative) number as a function of energy (we take the Fermi energy as zero).  
(b) Zoom of the spin-split midgap state. (c) Zoom of the conduction band minima.
For clarity, we substitute the delta functions composing the DOS by gaussian functions with a finite broadening.}
\end{figure}

We first revisit the single void samples. The ground state
of structures with single atom vacancies have one
unpaired electron within the $U=0$ model and, according to Lieb's theorem,
spin one half in the finite $U$ model. Our mean field
calculation, for $U=2$eV agrees with the Lieb theorem.
There is a spin splitting of the midgap state $\Delta_S$ and
a smaller spin splitting $\delta$ of the conduction and valence band states, as
shown in  Fig.  \ref{DOS}. The spin degeneracy is thus broken, with only
one of the spin channels of the midgap state occupied, the other being empty. This
results in a finite magnetization density, localized around the vacancy, as
shown in Fig.  \ref{figsingle}. Although the magnetization resides mostly in
the majority sublattice, interactions induce some reversed magnetization in the other
sublattice.

\subsubsection{Analytical model}
We can gain some insight by doing an analytical description of the mean field
results that involves some approximations valid when $U$ is
much smaller than the band-gap of the ideal ribbon $E_g$. In this case,
we assume that only the midgap state is spin polarized: 
 \begin{equation}
\langle m_i \rangle_0= \frac{1}{2}  |\phi_v(i)|^2
%\left( \langle n_{0\uparrow}\rangle -\langle n_{0\downarrow}\rangle \right)
 \end{equation}
where $|\phi_v(i)|^2$ is the $U=0$ square modulus of the midgap wave function. 
%At zero temperature the expectation value is $1$ ($0$) for the lowest (highest)
%energy state.
Notice that the normalization of $\phi_v(i)$ ensures that
the total spin of the ground state is consistent with Lieb's theorem, 
 $\sum_i \langle m_i \rangle_0 =\frac{1}{2}$.
The corresponding exchange splitting is
\begin{equation}
\Delta_S=\epsilon_{0\uparrow}-\epsilon_{0\downarrow}=
U \sum_i |\phi_v(i)|^2 \langle m_i \rangle_0 
=\frac{U}{2} \sum_i  |\phi_v(i)|^4
\label{lineargap}
\end{equation}

\begin{figure}
[hbt]
\includegraphics[width=\linewidth]{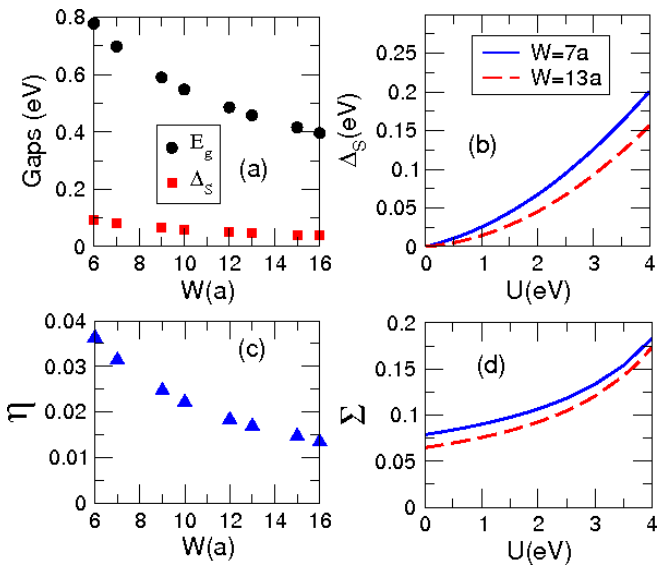}
\caption{ \label{gaps} (Color online) (a) Non-interacting ($U=0$)
energy gap (circles) and interacting ($U=2eV$) midgap spin
splitting (squares) as a function of the ribbon width $W$. 
 (b) Midgap spin splitting as a function of $U$ for two ribbon widths $W=7a$ and
 $W=13a$.  (c) $U=0$ inverse participation ratio for the midgap state.
 (d) Standar deviation of the magnetization $\Sigma$, as defined in
 Eq. \ref{sigmaJFR}, as a function of $U$ for two ribbon widths.  }
\end{figure} 

We see that, within the simplified analytical model, the
spin splitting of the midgap state is proportional to the inverse
participation ratio, $\eta = \sum_i |\phi_v(i)|^4$
This quantity measures the degree of localization of the zero-energy state.
An extended state in which the wave function is equally shared by $N$ atoms.
has  $\eta = \frac{1}{N^2}$. In the opposite limit where the state is localized
in a single atom we would have  $\eta = 1$. 
The inverse participation ratio shown in Fig. \ref{gaps}(c) corresponds
to a number of atoms in the range $N=5$ to $N=9$. 
As discussed above,
the localization of the midgap states plays an important role in the 
minimal distance at which they are effectively decoupled.

In Fig.   \ref{gaps}(a) we plot the $U=0$ gap of the ideal ribbon $E_g$ and the
$U=2$eV spin splitting $\Delta_S$ of the mid-gap  state 
as a function of the ribbon width $W$, as obtained from the full numerical
calculation. As discussed above,
we exclude the widths that give $E_g=0$. We see that the midgap spin splitting is a
decreasing function of $W$. This is related to the fact that, in the small $U$
limit, $\Delta_S$ is proportional to the inverse participation ratio $\eta$, which is also
a decreasing function of the ribbon width, as shown in Fig. \ref{gaps}(c).  
The extension
of the midgap state increases as the ribbon becomes wider, resulting in a
reduction of the midgap spin splitting. As $E_g$ tends to zero (bulk graphene) the midgap state becomes
non-normalizable\cite{pereira06} and $\Delta_S$ is expected to vanish (see below). 

Whereas the total magnetic moment $\sum_i \langle m_i \rangle$ is given by the sublattice
imbalance, the degree of localization of the spin texture is not.
In order to quantify it we define the standard deviation
\begin{equation}
\Sigma=\sqrt{\sum_i \langle m_i \rangle ^2}.
\label{sigmaJFR}
\end{equation}
In this definition $\Sigma$ is not normalized as usual by $N$, the total number of
atoms of the sample, since $\Sigma$ characterizes a localized object. For
sufficiently large simulation cells, doubling $N$ would imply a decrease of a
normalized $\Sigma$ without changing the local properties of the localized
 magnetic texture. 
Notice that, within the analytical model valid for $U<<E_g$, we have $\Sigma
\simeq \frac{1}{2} \sqrt{\eta}$. Hence, in the absence of staggered magnetization,
both $\Sigma$ and $\eta$ would measure the localization of the magnetic moments.
For instance, if $\langle m_i\rangle >0$ at all sites and  
 $S=1/2$, the maximal $\Sigma$ would be 0.5. However, 
 the graphene lattice
 responds with a  staggered magnetization to the presence of defects and
 $\Sigma$ {\em also} measures the magnitude of that response.  
  In Figs. \ref{gaps}(b) and (d) we plot the mid-gap spin splitting $\Delta_S$ 
   and $\Sigma$ for two ribbons with $W=7a$ and $W=13a$  as a function of $U$. The midgap spin
splittings can be fitted to
$\Delta_S(N_y=7)(U)=0.0172 U + 0.0082 U^2$ and to
$\Delta_S(N_y=13)(U)=0.0062 U + 0.00821676 U^2$. 
According to Eq. \ref{lineargap} the linear coefficients 
should be compared to $0.5\eta$, which is 0.016 for
$N_y=7$ and $0.008$ for $N_y=13$. The non-linear terms  arise from 
the interaction-driven mixing
between the midgap states and the conduction states. This is also consistent
with the fact that $\Sigma$ increases as a function of $U$ as shown in the lower
panel of Fig. \ref{gaps}(d). The staggered magnetization is an increasing function of $U$. 
The coefficient of the cuadratic term decreases with
the length of the sample, namely,  with the distance between
vacancies since we are using periodic boundary conditions. 
We will come back to this issue in the last section. 

\subsubsection{Spin-charge separation for $U>0$}

\begin{figure}
[hbt]
\includegraphics[width=\linewidth]{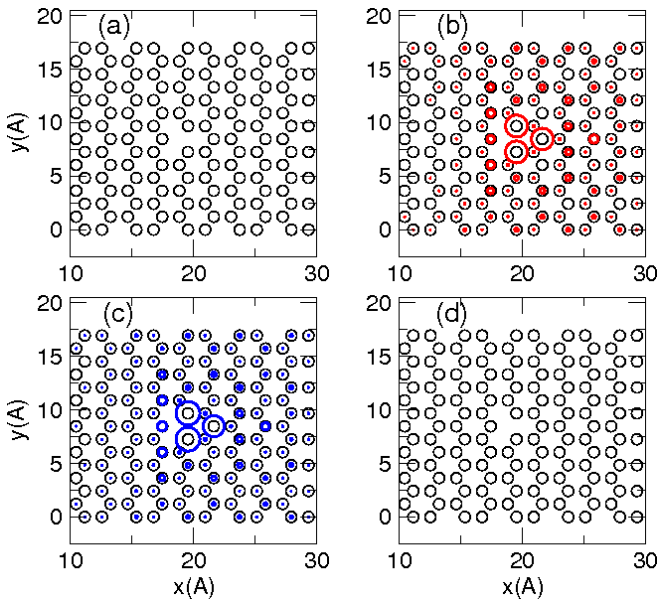}
\caption{ \label{spin-charge-sep}(Color online) 
Spin charge separation in single atom vacancy. Left column: neutral case. Right
column: Charged case. Upper panels: Charge density $q_i-1$. 
Lower panels: $|\langle m_i \rangle|/\sum_i |\langle m_i \rangle|$. The local charge and local spin are zero
everywhere for the neutral and charged cases respectively. The spin texture of
the neutral case is identical to the charge texture of the charged case. }
\end{figure}

We have verified that the ground state of structures
with  single atom vacancies 
 are locally neutral also with $U>0$: The integrated electronic
occupation in every site is one.  Hence, a localized 
spin texture with  total spin $1/2$ occurs in
the absence of any charge localization.  Our numerical results show that the
addition of an extra electron to single vacancy structures
results in a many electron state with total spin $S=0$,  local magnetization
which is zero everywhere, and local charge accumulated in the same atoms and with the
same distribution than the magnetic texture of the charge neutral structure. 
These results are shown in Fig. \ref{spin-charge-sep} for a ribbon with
$W=7a$ and $U=2eV$. Hence, it is apparent
that  the monoatomic vacancy results in a multielectronic state with spin charge
separation: The neutral ground state has a net electric charge $q=0$, but a total spin
$S=1/2$ localized in  a non-homogeneous spin texture in  locally neutral
atoms. The charged ground state has a net charge $q=-1$,  total spin $S=0$, no
local magnetic moments, and a  charge texture localized at the same location
than the spin texture of the neutral ground state.  This phenomenon resembles that
reported by Su-Schrieffer-Heeger\cite{SSH} in polyacetylene.  
%It would be interesting to see how this is related to the
%fractionalization mechanisms discussed in recent work in relation with textured
%distorsions of the hopping matrix\cite{Mudry-07,Jackiw-07,Dung-07,Herbut-07}.

\subsubsection{Larger voids}

\begin{figure}
[hbt]
\includegraphics[width=\linewidth]{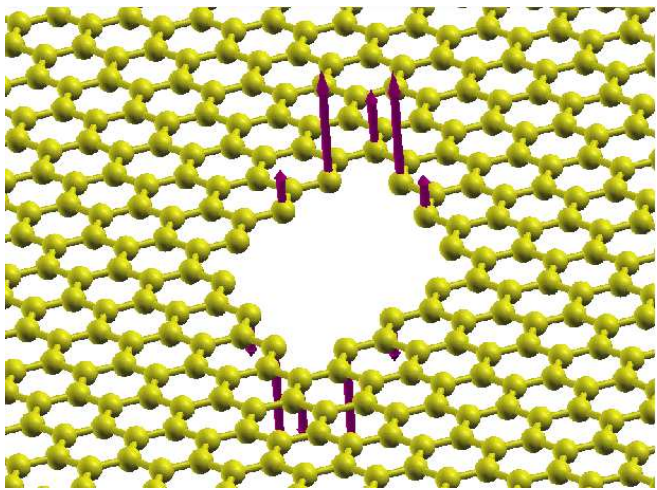}
\caption{ \label{rombo} (Color online) Emerging ferrimagnetic order in a rhombohedral void with
 imbalance charge $N_I=3-3=0$ situated the middle of a ribbon with $W=10a$i for
 $U=2$eV. The largest magnetic moment per atom is $\langle m_i \rangle=0.05$ }
\end{figure}

We have also calculated the mean field magnetic structure for sublattice imbalanced
larger voids. In Fig. \ref{figsingle}(b) we show the magnetization profile for
a triangular void with $N_I=2$.  For the chosen
value of $U=2$ eV the staggered magnetization is barely visible in this scale.
In agreement with Lieb's theorem, it has a spin $S=1$, made out of local moments localized, mostly, on
the triangle boundaries. This object is the somehow complementary of the
triangular graphene islands considered recently by two of us\cite{JFR07}. 
Figure \ref{rombo} shows the ferrimagnetic
spin texture around a rhomboidal void with imbalance charge $N_I=3-3=0$, 
i.e., composed of two triangular voids with $N_I=\pm3$. Local moments 
with $\langle m_i\rangle \simeq 0.05$, three times smaller than those formed in
the edges of infinite length zigzag ribbons,  are formed on opposite corners of the
void. We have verified that, for $U=2$ eV, the smallest void of
this shape which features local moments is the one of the figure.  The
rhomboidal void is similar to the hexagonal islands considered in Ref. \onlinecite{JFR07}
in the sense that both have $S=0$ and develop local moments if they are sufficiently large.

\subsection{Two vacancies with $U\neq 0$}
We now study the interaction between two magnetic defects with local
sublattice imbalance $N_I=\pm 1$. The Lieb's theorem warrants that, when the sign
of the sublattice imbalance is the same for the two defects, the total spin of the
ground state is the sum of the spin of the individual defects. Hence, they are
coupled {\em ferromagnetically}\cite{Kumazaki07,Yazyev07,Pisani07}.
In contrast, if the  two defects have opposite
sublattice imbalance so that the global sublattice imbalance is zero, Lieb's
theorem warrants that the total spin is zero. Our calculations show
that this can happen in two different scenarios: The local magnetization
might be zero everywhere or the two defects could be magnetized
along opposite directions, i.e., could be coupled {\em
antiferromagnetically}. When the defects are sufficiently far apart from each other their local
electronic structure should be identical to that of single defects. 
 Hence, the
spin interaction between two magnetic defects can be either ferromagnetic or
antiferromagnetic, as in the case of indirect exchange interactions (RKKY) between
single site magnetic moments\cite{RKKY-Luis,Saremi},
but   can also result in the annihilation of the local magnetic order, an
scenario that goes beyond the RKKY picture. 

\begin{figure}
[hbt]
\includegraphics[width=\linewidth]{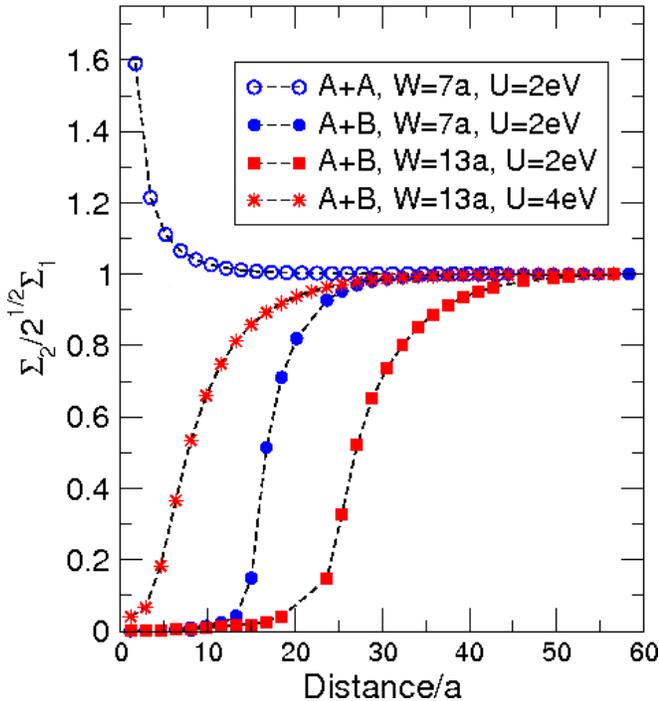}
\caption{ \label{pairs} (Color online)
 Normalized magnetization as a function of the
distance between vacancies for four cases. 
 Ribbon with $W=7a$, $U=2$eV,  vacancies in the same
sublattice (open circles), same ribbon, 
holes in different sublattices (full circles),
ribbont with $W=13a$, holes in different sublattices, $U=2$eV (full squares) and
$U=4$eV (stars)}
\end{figure}

In Fig. \ref{pairs} we plot the normalized standard deviation of the
two magnetic moments $\Sigma_2$  for ribbons with $W=7a$  and $W=13a$  as a function
of the defect separation.  We normalize the computed $\Sigma_2$ to the one 
corresponding to two independent single-defect magnetic textures,   
$\sqrt{2}\Sigma_1$. Here  $\Sigma_1$ is the computed single vacancy standard
deviation  in the same ribbon.  When
the defects are sufficiently far away $\Sigma_2$ must tend to
 $\sqrt{2}\Sigma_1$, i.e., the normalized $\Sigma_2$ must tend  to $1$.
We consider the effect of the ribbon width
$W$, interaction strength $U$, and sublattice imbalance upon the magnetic
interactions between the two defects.   In the case of $W=7a$
we show both monoatomic vacancies lying on the same sublattice (A+A, open
circles), whose ground state total spin is $S=1$, and
on different sublattices (A+B, full circles), whose ground state spin is
$S=0$.  The two curves for $W=7a$ are calculated with $U=2$eV.  At large distances,
the two defects become decoupled, as expected. At short distances, the behaviour
of the magnetic texture is radically different for both A+A and A+B structures. In
the former case $\Sigma$ is enhanced, indicating the localization of the
magnetic texture in a smaller region. Since the total spin is $1$, local moments
survive even when the two defects are very close.
As the separation between defects increases they become
independent from each other and  $\Sigma_2=\sqrt{2}\Sigma_1$. When this
happens, the energy gap between $S=1$ and $S=0$ should vanish. This is an example of rule number 1. 

In contrast
to the A+A case, the {\em local} magnetization of the  A+B structure  vanishes
below a  {\em minimal distance} $D_c$. This is an important result.
In other words, there is a {\em maximal density} of
defects above which zero energy states hybridize and local magnetic moments
vanish.  The critical density depends on the energy scales of the problem, the single
particle gap $E_g$, controlled by the ribbon width, and the on-site repulsion
$U$. For fixed $U$ the decoupling distance is definitely shorter
for $W=7a$ than for $W=13a$. Hence, the critical (linear) density becomes {\em smaller},
as the ribbon width {\em increases}.  This is consistent with the fact that both
the $U=0$ inverse participation ratio and the $U\neq 0$ $\Sigma_1$ are
decreasing functions of the ribbon $W$. The wider the ribbon, the larger the
delocalization of the zero energy state. Hence, the hybridization between the
midgap states associated with each vacancy survives at a larger distance for
wider ribbons. Finally, in fig. \ref{pairs} we also show $\Sigma_2$
for $W=13a$ and two values of $U$, 2 and 4 eV.
The decoupling distance
(critical density) decreases (increases) as a function of $U$. In other words,
interactions drive the system magnetic, as expected. The A+B case is the simplest example that exemplifies
rules number 2 and 3.

According to the two-vacancy calculation shown in the figure,  the magnetic (low
density, large inter-vacancy distance) and non-magnetic (high density, low
intervacancy distance) phases are separated by a crossover region.  If we take
as an estimate of the critical distance below which local moments are quenched
the distance for which  $\Sigma_2(D_c)/\sqrt{2}\Sigma_1=0.5$, we find that 
$D_c=40\AA$ for $W=7a$ and $D_c=65 \AA$ for $W=7a$, both for $U=2$eV. The
corresponding  critical {\em linear} densities $n_c\equiv \frac{1}{D_c}$ are 
$n_{1c}(W=7a)=2.5 \times 10^{6} $cm$^{-1}$ and 
  $n_{1c}(W=13a)=1.5 \times 10^6 $cm$^{-1}$,
respectively. The corresponding areal densities, $n_{2c}=\frac{1}{W
\times D_c}$ are $n_{2c}(W=7a)=1.4 \times10^{13} $cm$^{-2}$ and 
$n_{2c}(W=13a)=4.8\times
10^{12} $cm$^{-2}$ respectively. These numbers should be taken as order of
magnitude estimates of the real critical density. 

In the case of A+B pairs, the crossover from the locally magnetic to the
non-magnetic state is similar to the one described in compensated graphene
nanoislands\cite{JFR07}: Small islands are non-magnetic and larger islands have
magnetic edges.  
The critical density depends on the extension of the magnetization, which
in turn, depends on the ribbon single particle gap $E_g$ (which controls the extension of
the $U=0$ midgap states) and on the on-site repulsion $U$.
The quenching of the local moments in the A+B structures
is definitely related to the hybridization of the mid-gap states described in the
non-interacting model. This phenomenon has an analog in zigzag ribbons.
The  midgap bands are linear combinations of top and bottom edge states. 
The hybridization is negligible in the Brilloin zone boundary , and is much
larger in the middle. As a result, the exchange interaction strongly renormalize
the zone-boundary states, opening a magnetic gap, but they barely change in the
middle of the zone\cite{JFR-preprint}. 

\subsection{Defective graphene ribbons as diluted magnetic semiconductors}

The physical picture that emerges from the previous discussion leads to an
interesting conclusion: A semiconducting graphene ribbon with a density of
vacancies that induce magnetism will behave like a diluted (para-)magnetic
semiconductor (DMS)\cite{DMS} provided that the density of defects is smaller than the
critical density defined above (this would be an example of rule number 3). 
Charged excitations will present a gap and spin excitations will not. 
The long range ferromagnetic order found by Pisani {\em et al.} (see Ref. 
\onlinecite{Pisani07}) only occurs when the vacancies are all in
the same sublattice. 
It remains an open issue whether or not such a sublattice imbalance might occur in reality. 
Unless this can be shown, one should not expect long range ferromagnetic order in real samples.  

Interestingly, the conduction and valence bands depend
on the magnetic order of the local moments,
which might be induced by application of an external magnetic field.  
In the DMS case, the conduction and valence bands
are exchanged coupled to the local moments, provided by Mn atoms. At zero field
the Mn spins are randomly oriented and the average spin splitting of the bands
is zero. Application of an external field orders the Mn spins, resulting in a
finite average exchange induced spin splitting of the bands which is much larger
than the standard Zeeman splitting. This is known as giant Zeeman splitting.

\begin{figure}
[hbt]
\includegraphics[width=\linewidth]{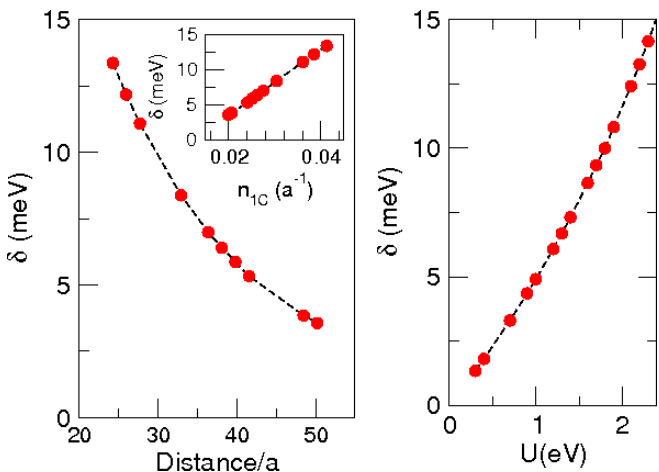}
\caption{ \label{figure-DMS} (Color online) Left panel: Bottom of the conduction band spin
splitting,  $\delta$, as a function of the defect distance and as a function of
the linear defect density (inset) for a ribbon with $W=7a$ and $U=$2eV. Right
panel: $\delta$ for the same ribbon for a fixed defect density as a function of
$U$.  }
\end{figure}

The same scenario  might occur in  semiconducting graphene ribbons with magnetic
vacancies. When the 
inter-defect distance is larger than the critical spacing,
$D_c$, the  local moments  are independent. 
Application of a magnetic field aligns them and
 induces an exchange-induced splitting of the conduction and valence
bands much larger than the intrinsic Zeeman splitting\cite{DMS}.  
In the case of graphene ribbons with vacancies
we have computed the
spin splitting of the bottom of the conduction band 
\begin{equation}
\delta\equiv E_{\uparrow}-E_{\downarrow},
\end{equation}
where $E_{\sigma}$ is the first level above the mid-gap state [see Fig.
\ref{DOS}(c)]. Notice that the shift of the top of the valence band and the
bottom of the conduction band is such that the gap, ignoring the midgap states,
is {\em spin-independent}.
Since we consider independent defects, the calculation is done with a single
defect per unit cell.
In Fig. \ref{figure-DMS} we plot $\delta$ for a ribbon with $W=7a$. 
In the left panel we plot $\delta$ as a function of the inter-defect distance
considering only values bigger than $D_c$ for which the defects are decoupled.
For a $U=2$eV we find that $\delta$ ranges between 5 and 15 meV. This
splitting could be obtained with an applied magnetic field such that $g\mu_B B>>
kT$,  yet, $\delta>>g \mu_B B$.  As in the case of real DMS, this giant
Zeeman splitting scales linearly with the defect density, as shown in the inset
of the right panel of Fig. \ref{figure-DMS}. Since there is a maximal
density above which the local moments are coupled and eventually they vanish,
the splitting $\delta$ can not be increased indefinitely. This phenomenon also
has an analog in
DMS: Direct antiferromagnetic coupling between Mn spins eventually blocks the
paramagnetic coupling to the external field. 

In the right panel of Fig. \ref{figure-DMS} we plot $\delta$ as a function of
$U$. For small $U$ we find that $\delta$ is almost linear
with $U$. This can be understood in the framework of the analytical model
discussed above. If we consider that, to lowest order, the magnetization only
comes from the mid-gap states with wave function $\phi_v(i)$ and we compute the splitting of the
bottom of the conduction band states to first order perturbation theory, we
obtain:
\begin{equation}
\delta=U \sum_i |\phi_c(i)|^2 \sum_{v}|\phi_v(i)|^2
\end{equation} 
where $\phi_c(i)$ is the $U=0$ single particle state of the bottom of the
conduction band. If we approximate $\phi_c\simeq 1/\sqrt{N}$, where $N$ is the
number of atoms in the unit cell, and we use
the normallization condition of the midgap states, $\sum_i |\phi_v(i)|^2=1$ 
then we have:
\begin{equation}
\delta\simeq \frac{UN_v}{N} 
\end{equation} 
where $N_v$ is the number of magnetic vacancies per unit  cell.
This equation accounts also for the fact that $\delta$ scales linearly with the
defect density. Deviations from the linear behaviour arise due to the
magnetization that arises from states other than mid-gap states.

The strong sensitivity of the conduction states of the defective armchair
ribbon on the application of a moderate magnetic field
should give rise to strong spin-dependent magnetotransport and magnetooptic 
effects, in analogy with DMS spintronic devices. Notice that, in contrast
to standard Mn doped II-VI semiconductors, for which electrical injection of
carriers results in a new carrier mediated coupling\cite{Boukari,JFR04},
the addition of carriers in
this system would results in the compensation of the mid-gap states and the
disappearance of the local moments, as shown in Fig. \ref{spin-charge-sep}.

%Finally, we present results for a larger triangular void ($|N_I|=2$). This
%presents magnetization along the inner edge [see Fig. \ref{single}(b)] which is
%where the zero-energy states are localized [see inset in Fig. \ref{single}(b)].
%As for vacancies,  as the width of the ribbon increases and  $\Delta_{\rm
%gap}\l$ decreases the magnetization spreads over more lattice sites.

\section{Defects in bulk graphene: vacancies}
\label{bulk}

So far we have considered the electronic structure of semiconducting graphene
ribbons with vacancies and voids. As shown in Fig. \ref{pairs} the critical distance for the quenching of the magnetic 
moments increases with the ribbon width. An important question is whether or not this critical
distance converges to a finite value in the two-dimensional limit.
We have also seen in Fig. \ref{gaps}(d) how the (standard deviation of the) magnetization $\Sigma$ associated
with vacancies decreases as the gap of the ribbon decreases.
In this section we address the
question of what happens to these and other results obtained above in the limit of infinitely wide ribbons where the gap
goes to zero, i.e., bulk graphene. 
The extrapolation to the two-dimensional case is not straightforward.
We thus consider a new strategy. Here we consider unit cells with periodic boundary conditions in both directions.
An infinite graphene crystal
with a unit cell formed by $N_y$ parallel armchair-like chains, each of them
containing $N_x$ carbon atoms. 
%Notice that the role of $N_x$ and $N_y$ are
%interchanged with respect to the previous section.
The dimension of this unit cell is $(N_x 
\frac {\sqrt{3}}{4} a,N_y a)$ being $a$ the graphene lattice parameter. We are
interested in square unit cells and therefore we consider units cells
$(N_x,N_y)$ of sizes (24,10), (32,13),(40,16), (48,20), (60,26), and (72,31). We
locate one or more vacancies in the unit cell considered and, as for ribbons, we
obtain the eigenvalues, eigenfunctions, and the magnetization
at each place of the system, $\langle m(i)\rangle $, by solving self-consistently the
Hamiltonian. 
%It is important
%to note that in this work we only consider undoped graphene, and
%because the electron-hole symmetry, the electronic charge at
%everyplace is always the unity even in the presence of vacancies.
%The local charge neutrality allows us to ignore the Hartree term of
%the long range Coulomb interaction\cite{JFR-ribbonAC-07},
%and partially justify the use of
%the Hubbard model for describing the electron-electron interaction.

\subsection{System with vacancies of the same type}
We locate a vacancy at the center of the unit cell in such a way
that our system describes  a square lattice of vacancies all located
on the same type of atoms, e.g., B in our notation. We obtain that, in
agreement with Lieb's theorem, the ground state of the system has
a magnetic moment $S=1/2$ per vacancy. 
We also obtain the spin gap of the system $\Delta_S$, the wavefunction
$\phi _v (i) $ of the first empty state, and the local
magnetization.  In the limit of a large unit cell
$\Delta_S /2 $ and $\phi _v (i)$  should be the energy and the
wavefunction, respectively, of an isolated vacancy. In two dimensions we can characterize
the linear size  of the vacancy wavefunction by its first moment
\begin{equation}
\langle R \rangle  = \sum _{i } |\mathbf{r}_i - \mathbf{r}_0| |\phi_v (i)|^2 \, \,  \, \, \, ,
\end{equation}
where $\mathbf{r}_0$ is the position of the vacancy and the sum is
over all atoms of the unit cell. As discussed above, the bipartite character of the
graphene lattice produces an antiferromagnetic coupling between the
magnetization of the two sublattices of the
system\cite{RKKY-Luis,Saremi}. We define the linear size of the
magnetization in each  sublattice as
\begin{equation}
M_d ^{A(B)}  = \sum _{i \in A(B) } |\mathbf{r}_i - \mathbf{r}_0| \langle m_i\rangle
\, \, \, \, \, .
\label{sizemag}
\end{equation}
$M_d ^{A}$ and $M_d ^{B}$ have opposite signs, and their sum, $M_d =
M_d ^{A} +M_d ^{B}$ indicates the extension of the net magnetization.

\begin{figure}
  \includegraphics[clip,width=9cm]{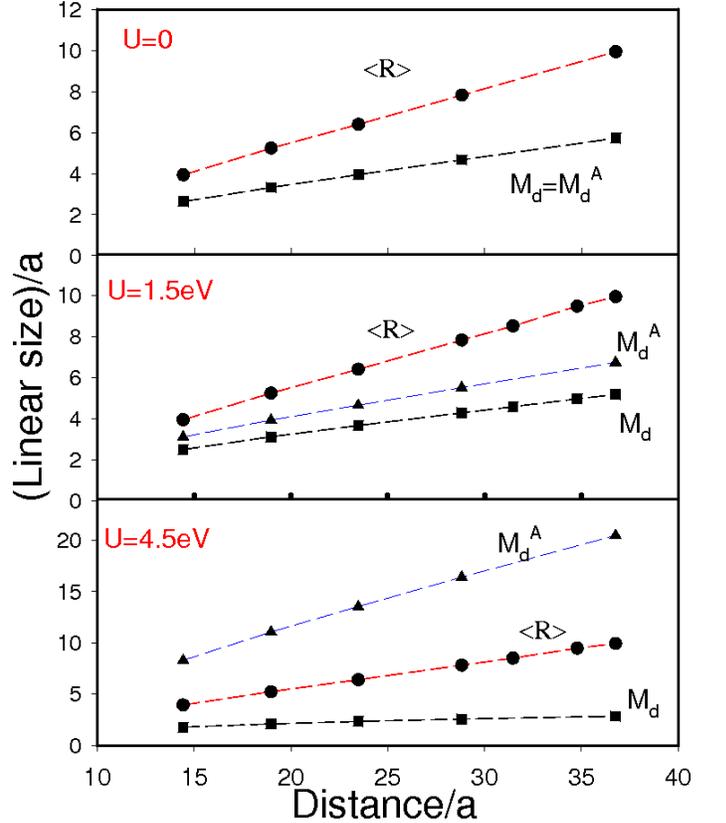}
  \caption{(Color online) Linear size of the wavefunction of the
  vacancy, $<R>$, of the net magnetization, $M_d$,
  and of magnetization in the sublattice A, $M_d ^ A$, as function of the distance between
  vacancies. From top to bottom the panels  correspond to Hubbard
  constant,
  $U=0$, $U=1.5$ and $U=4.5$ eV.
  The vacancies are located in sites type B.  }
   \label{size}
\end{figure}

In Fig. \ref{size} we plot the linear size of the vacancy
wavefunction and of the magnetization as function of the distance
between vacancies for different values of $U$. The first thing to
note is that the size of the wavefunction increases linearly with
the size of the unit cell, practically independent of the value of
$U$. This result indicates that the electron-electron interaction
almost does not affect $\phi _v$ which,  when $U=0$, becomes a
quasilocalized state with weight in only one sublattice (A) and decays
as $1/r$, in agreeement with analytical results\cite{pereira06}.
When $U=0$ only the sublattice A is
magnetized and $M_d$=$M_d^A$. However, $M_d$ is considerable smaller
than $\langle R\rangle $, indicating that the presence of the vacancy does not just
creates a quasilocalized state, but also modifies strongly the
wavefunctions of the states in the continuum. 
Notice the difference with the ribbons where
the magnetization follows the wavefunction for sizable confinement gaps.

As we increase $U$, the magnetic texture  evolves in such a way
that their size $M_d$, as defined in Eq. (\ref{sizemag}), decreases. 
This is accompained by an increase of the staggered magnetization, reflecting 
 the antiferromagnetic tendency of the bipartite lattice, which
 polarizes
the sublattice B in the opposite direction that sublattice  A with $M_d ^A > M_d $.
This effect can be rather dramatic for moderate values of $U$, in
Fig. \ref{size} we show that for $U=4.5 $ eV the extension of the
magnetization in  sublattice A is considerable larger than $\langle R \rangle$
and $M_d$. Note that $U= 4.5$ eV is still below the critical value
$U_{AF} \sim 5.5$ eV for the occurrence of an antiferromagnetic
instability in perfect graphene\cite{Sorella92,Peres04,Kumazaki07}. Thus, a
network of vacancies in the same sublattice would have a magnetic
ground state, in agreement with Lieb theorem, and enhanced staggered
magnetization, compared to perfect graphene.

\begin{figure}
  \includegraphics[clip,width=9cm]{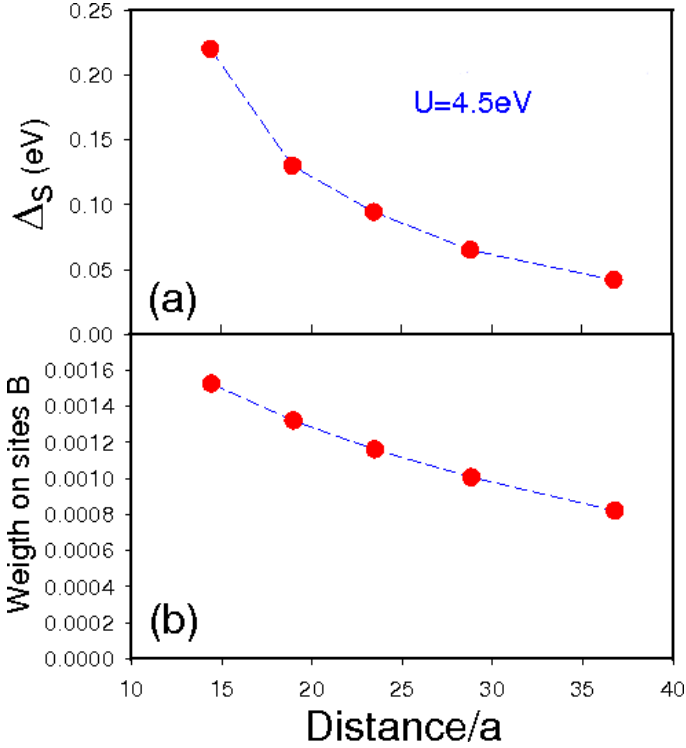}
  \caption{(Color online) (a) Midgap spin splitting as function of the
  distance between vacancies and (b) weight of the quasilocalized wavefunction on
  sites B. The vacancies are located in the sublattice  B and $U=4.5$ eV.  }
   \label{gap}
\end{figure}

We now consider the midgap spin splitting $\Delta_S$ in  two dimensional graphene with
a finite density of vacancies  in the same sublattice. 
%In the single vacancy case $\Delta_S$  is originated by the Hubbard term.
In the previous case of semiconducting ribbons there was a
strong indication that $E_g > \Delta_S$ for any ribbon width in the single
vacancy limit. Hence, we might expect that $\Delta_S$ vanishes in two dimensional graphene.
When we have a finite density of defects, $\Delta_S$ has also an inter-defect 
contribution arising from the hopping term. 
This mechansim is 
possible only if the midgap states have weight on the two sublattices.
Midgap states associated with defects 
in the same sublattice can only be coupled 
through interaction driven sublattice mixing.
As the gap is a product of the coupling between the magnetizations induced by
the Hubbard interaction, the value of the gap increases quadratically with $U$
to lowest order.
Our calculations in Fig. \ref{gap}(a) show that
for  $U \ne 0$, the midgap spin splitting 
$\Delta_S$ goes to zero as the density of vacancies goes to zero. 
In Fig. \ref{gap}(b) we plot the weight of $\phi _v (i)$ on the
sublattice where the vacancy is located.
This quantity also tends to zero as the density of defects decreases.
Thus, in the single impurity limit the spin gap goes
to zero and the interacting $\phi _v (i)$ lives only on one sublattice. 
%In this
%limit, although the electron-electron interaction creates a local
%staggered magnetization near the vacancies, $\phi _v (i)$ has
%only weight in one sublattice of the crystal.

%As in the case of semiconducting ribbons, 
%vacancies located in the same sublattice are not coupled by  the
%kinetic energy and, in absence of electron-electron interactions, they have zero
%energy. For $U \ne 0$ a staggered magnetization near the vacancies appears and
%vacancies located on the  same lattice couple. This interaction  opens an
%energy  gap for the two spin orientations of the wavefunction localized at the
%vacancy. This gap decreases with the distance (see Fig. \ref{gap}(a)) and in the
%diluted limit the energies  of the vacancies are degenerate at the Fermi energy.

\subsection{Vacancies on different sublattices: Compensated case} 
In principle, one could expect that in real graphene samples the number of vacancies on
sublattices A and B are roughly equal. 
In this situation  Lieb's
theorem requires  that the total spin of the system should be essentially zero.
From the results in the case of ribbons one should expect an antiferromagnetic
coupling between the magnetic moments for small concentrations of
vacancies. For large concentrations the local magnetic moments should disappear
and the sample should turn non-magnetic. In order to study the interaction
between vacancies located in different sublattices and the local magnetization
in a compensated  system, we locate two vacancies with total
sublattice imbalance  equal to zero in the unit cell. 
The A and B impurities form
two interpenetrated square lattices in such a way that the distance
between impurities is maximum. We have checked that in agreement
with Lieb's theorem the $S$=0 solution is the ground state of the
system.

\begin{figure}
  \includegraphics[clip,width=9cm]{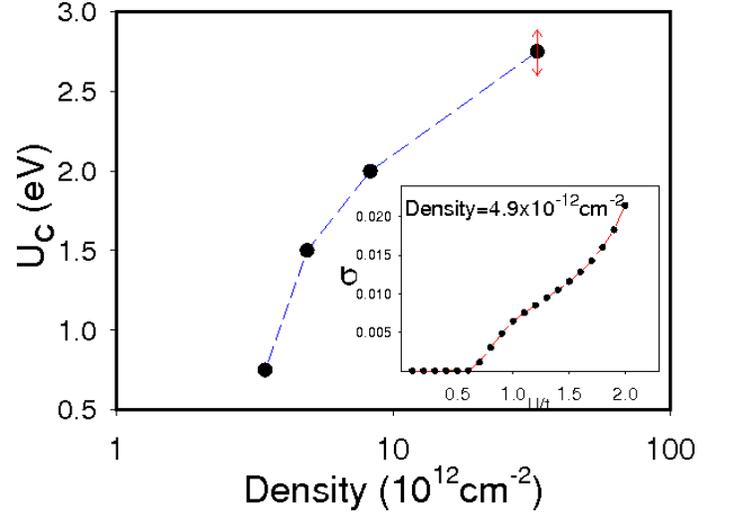}
  \caption{(Color online) Value of the critical values for the
  occurrence of a magnetic texture in a system of vacancies on atoms of type
  $A$ and $B$  forming two square lattice interprenetrated lattices.
  The inset indicates the variation of the standard deviation of the
  local magnetizations as function of the Hubbard constant for a
  system of dimension (26,60). The error bar is an estimation of the
  numerical error in the calculations.
  }
   \label{U_cri}
\end{figure}

We quantify the local magnetization studying the standard
deviation of $\langle m_i \rangle $,
\begin{equation}
\sigma = \sqrt{\frac {\sum _{ i} \langle m _i\rangle  ^2}{N}} \, \, \, ,
\end{equation}
where the sum is over all carbon atoms and $N$=$N_x \times N_y$
is the number of atoms in the unit cell. In the inset of Fig. \ref{U_cri} we plot
$\sigma$ as a function of $U$ for a unit cell of size
$(26,60)$, which  corresponds to a density of vacancies of $0.5 \times
10 ^{13}$ cm $^{-2}$. We obtain that for small values of $U$ the
magnetization is zero everywhere and that there is a critical value
of the Hubbard coupling, $U_c$,  for which a local magnetization
near the vacancies appears\cite{Kumazaki07}. This critical $U$ depends
on the density of vacancies, and in Fig. \ref{U_cri} we plot $U_c$ as
function of the density of vacancies.  We obtain that $U_c$
decreases with the density of vacancies and from our results we
conclude that the in the limit of zero density $U_c$ tends to zero.
In that limit, each vacancy hosts a spin one half texture, which is decoupled
from the others.  
For high density of vacancies and moderate values of $U$, the
kinetic energy coupling between the vacancies is stronger than the
electronic repulsion, and in order to minimize the energy the system
makes the local magnetization zero everywhere. When the density of
impurities decreases, the kinetic energy coupling between vacancies
decreases and eventually the impurities become uncoupled and  each
of them gets a  total spin $\pm 1/2$ and the system behaves as a
diluted antiferromagnetic system\cite{RKKY-Luis}.
Unless a reason for global lattice imbalance exists (unknown to us),
for realistic values of $U$ in the range $1<U<2$ eV\cite{JFR07} we expect
that in highly irradiated graphene samples the local magnetization should be
zero across the sample. %The zero density  limit should not be confused with pure graphene, for
%which the mean field Hubbard model gives an antiferromagnetic phase for $U=2.2
%t$\cite{Sorella92,Peres04}. 

\section{Discussion and conclusions}
\label{conclusions}

Some of the possible limitations of our approach have already been mentioned in the beginning.
We have left aside the issue of
the structural stability of passivated vacancies and voids. 
Away from the edges,  a single monoatomic
vacancy might not result in a local atomic configuration that can be described
with our model. On the other side, the effect of a hydrogen atom atop a carbon
atom on graphene is very similar to the one described by our
model\cite{Yazyev07}. Hence,
the anomalous magnetic behaviour of irradiated graphite might be related to
$H-C$ pairs, rather than to missing atoms. It is also important to signal that
the mean field  approximation is known to overestimate the appearence of magnetic
order and yield critical values of $U/t$ smaller than those obtained with
methods that include quantum fluctuations\cite{Sorella92}. Finally, we have
neglected both second neighbor hopping and interatomic Coulomb repulsions.
Interestingly, both DFT and mean field Hubbard model
yield a very similar  description of the 
magnetic behaviour of graphene islands\cite{JFR07} and zigzag graphene ribbons\cite{JFR-preprint}. 
This indicates that the couplings neglected in the simple
Hubbard model have a small effect on the low energy electronic structure that
dominates the physical properties. 

We now summarize the main conclusions of this work. In the context of our model, the  main results are:
\begin{enumerate}

\item The electronic properties of the defects arising from  the removal of
atoms from graphene depend dramatically on an integer number, the sublattice
imbalance (or imbalance charge) $N_I=N_A-N_B$, which counts the difference  in the total number of
atoms per sublattice removed from a perfectly balanced graphene
lattice. $N_I$ can take values $0,\pm1, \pm 2$ etc.

\item It can be rigorously shown\cite{Abrahams} that the single-particle
spectrum of a structure with sublattice imbalance $N_I$ has, at least, $|N_I|$
midgap states per spin channel, occupied by $|N_I|$ electrons in neutral
graphene.

\item Repulsive Coulomb interactions will result in a many-body ground state
with $2S=|N_I|$. This is an exact result in the case of the Hubbard model\cite{Lieb89}. 
%The spin of the ground state 
%of a structure  is  related to the sublattie imbalance through $2S=|N_I|$.

\item Whereas the global electronic structure of a 
given graphene system is given by the Lieb, the local structure is not. By assigning
 {\em local} sublattice imbalance numbers to defects, provided that they are sufficiently apart, a set of 
rules to predict basic features of the magnetic structure has been proposed. 

\item We find that single voids with $|N_I=1|$ give rise to states with
 spin-charge separation, in the sense that a
localized magnetic texture does not entail charge localization. The addition of
a single electron to the system 
results in a many-body state with $S=0$ and the disappearance of the magnetic
texture, which is substituted by a  charge texture, as seen
in Fig. \ref{spin-charge-sep}. In this sense, 
the properties of these states are very
similar to Su-Schrieffer-Heeger midgap states\cite{SSH}.

\item The addition rules for two voids with a givel local
sublattice imbalance or imbalance charge present similarities with those of vortices, e.g., in 
superconductors.  When sufficiently apart,
two voids with local imbalance $+N_I$ and $-N_I$ behave like two
independent objects with local spin $2S=|N_I|$. Below a certain
distance they annihilate each other and  the local magnetization vanishes
(Fig. \ref{pairs}). When two voids with the same sign are brought
together,  they result in a region with enhanced local magnetization and spin
$2S=|N_I|+|N_I|$, as seen in Fig. \ref{figsingle}.

\item In analogy with graphene nanoislands\cite{JFR07}, 
sufficiently large voids with $N_I=0$ can still have local magnetic
moments. These can interpreted as if the large void with $N_I=0$ was the sum of
two voids with  $\pm N'_I$. An example of this is the rhomboid of Fig. 
\ref{rombo}, obtained from merging two triangular voids with $N_I=\pm 3$ back to
back. 

\item Our results show that spin interactions between two magnetic defects of same $|N_I|$
can be of three types:
Ferromagnetic, antiferromagnetic or annihilating. In the first case, the ground
state spin is the sum of the spins of the magnetic defects when infinitely
apart, in the second case the spin is the difference between those two. In the
third  case both the total and the local spins are zero. Antiferromagnetic
and annihilating couplings occur in lattices without  global sublattice
imbalance, whereas ferromagnetic coupling requires global sublattice imbalance.

\item Our simulations show that, in balanced defective structures, there is a
maximal density of monoatomic vacancies that can sustain local moments. In the
case of ribbons,  this critical density depends on the ribbon width. The
critical density also depends on  $U$. A phase diagram for bulk graphene is provided
in Fig \ref{U_cri}.

\item   Depending on the density of
 vacancies, distributed randomly in the two sublattices, we distinguish three
 phases. In the very dilute limit the system is paramagnetic, with some common
 properties with II-VI diluted magnetic semiconductors. In the opposite high-density 
limit the local moments are annihilated.  The intermediate phase
 features antiferromagnetically coupled local moments. This would be a
 realization of the  so-called diluted antiferromagnet\cite{RKKY-Luis}
   
\item We predict giant Zeeman splitting in the case of semiconducting ribbons in
the dilute limit. Uppon application of an external magnetic field such that
$g\mu_B H>k_B T$, the magnetic moment of all the defects would point parallel
to  applied field. This would result in a interaction induced splitting in the
band states, much larger than the ordinary Zeeman splitting, as seen in Fig.
\ref{figure-DMS}. 

\item A ferromagnetic phase is not expected for defective graphene, unless
vacancies occur predominantly in one of the two sublattices. Such an unbalanced
 situation would require further explanation.

\item In the case of zero-gap graphene, 
we find that midgap states survive, even in the interacting case,
in the very dilute limit (Fig. \ref{gap}).  Since ideal graphene is a semimetal, the
thermodynamics properties of graphene might be dominated by this type of
defects.

\end{enumerate}

Upon completion of this manuscript, a related work by O. V. Yazyev (aXiv:0802.1735) 
has been reported. 

\section{Acknowledgements}
We acknowledge fruitful discussions with F. Guinea, B. Korgel, and P. L\'opez-Sancho.
This work has been financially supported by MEC-Spain under Grants Nos.
MAT2007-65487, MAT2006-03741, and CONSOLIDER CSD2007-00010, and by
Generalitat Valenciana under Grant No. ACOMP07/054.

%\bibliography{matcon}

\end{document}